\documentstyle[prl,aps,multicol,twoside]{revtex}
\renewcommand{\narrowtext}{\begin{multicols}{2} \global\columnwidth20.5pc}
\renewcommand{\widetext}{\end{multicols} \global\columnwidth42.5pc}

\begin{document}
\draft

\title{\Large \bf Delbr\"{u}ck scattering at energies
$140\div450$~MeV.}

\author{
\small SH.ZH.~AKHMADALIEV, G.YA.~KEZERASHVILI, S.G.~KLIMENKO, V.M.~MALYSHEV,\\
\small  A.L.~MASLENNIKOV, A.M.~MILOV,  A.I.~MILSTEIN, N.YU.~MUCHNOI,\\
\small A.I.~NAUMENKOV, \fbox{ V.S.~PANIN},  S.V.~PELEGANCHUK, V.G.~POPOV,\\
\small G.E.~POSPELOV, I.YA.~PROTOPOPOV,  L.V.~ROMANOV, A.G.~SHAMOV,\\
\small D.N.~SHATILOV, E.A.~SIMONOV, YU.A.~TIKHONOV }
\address{Budker Institute of Nuclear Physics,
 630090 Novosibirsk, Russia}
\maketitle

\begin{abstract}
The differential cross section of  Delbr\"{u}ck
scattering  is measured on a bismuth germanate  $\rm(Bi_4Ge_3O_{12})$
target  at photon energies \mbox{$140\div450$~MeV} and scattering angles
$2.6\div16.6$~mrad. A good agreement with the
theoretical results, obtained  exactly in a Coulomb field, is found.
\end{abstract}
\narrowtext
\newpage

\section{\bf Introduction}

Delbr\"{u}ck scattering~\cite{0} is a process,
in which the initial photon turns into a virtual electron-positron pair,
it is  scattered in a Coulomb field of a nucleus and then transforms into
the final photon (Fig.~1a). Thus, the final photon energy is equal to
the energy of the initial photon (elastic scattering).

The interest to the experimental study of Delbr\"uck scattering has the
following motivations.
First, it is one of nonlinear quantum electrodynamic processes
accessible at present time to direct observation. Other such process is
photon splitting in a  Coulomb field (Fig.~1b). For these processes the
contribution of higher orders of the perturbation theory with respect to the
parameter $Z\alpha$ ($Z|e|$ is
the charge of the nucleus, $\alpha=e^2=1/137$ is  the fine-structure
constant, $e$ is the electron charge, $\hbar=c=1$) at large
$Z$ essentially modifies the cross section. Therefore, the investigation of
these processes can be used as a good test of quantum electrodynamics in a
strong electromagnetic  field.
Second, Delbr\"{u}ck scattering is a background process to the nuclear Compton
scattering, which is an effective experimental tool to study
mesonic and nucleon internal degrees of freedom of nucleus~\cite{1}.

At present, four methods of  Delbr\"{u}ck scattering amplitude
calculation are used, possessing different areas of applicability~:

I) The amplitude is calculated in the lowest in $Z\alpha$ order of the
perturbation theory, but for an arbitrary
photon energy $\omega$ and  scattering angle $\theta$.
The review of numerous results, obtained in this
approximation, can be found in \cite{2,3}.
These results are applicable only at small Z, when the parameter
$Z\alpha\ll 1$.

 II) At high photon energies $\omega \gg m$ ($m$
is the electron mass) and small scattering angles $\theta\ll 1$ the
amplitude is obtained by summing in a definite approximation of the
Feynman diagrams with an arbitrary number of photons exchanged with a
Coulomb centre~\cite{4}.

III ) At $\omega \gg m $ and $\theta\ll 1$ it is possible to use also a
quasiclassical approach~\cite{5}, since in this case the momentum transfer
\mbox{$\Delta=\mid {\bf k}_2-{\bf k}_1\mid=\omega\theta$}
 (${\bf k}_1$ and ${\bf k}_2$ being the momenta of the initial and final
photons, respectively), and the
characteristic angular momentum  $l\sim \omega / \Delta=1 / \theta\gg 1$.
Numerically, the approaches II and III lead to the same results, as
they should, and show the significant difference between the cross section
calculated exactly in $Z\alpha$ and the cross section obtained
in the lowest order of the perturbation theory.

IV ) At $\omega \gg m $ and $\theta \sim 1$ the amplitude is calculated
exactly in $Z\alpha$ but neglecting the electron mass as compared to
$\omega$ and $\Delta$~\cite{6,7}. The approach is based on the
use of the relativistic electron Green function in a Coulomb
field. In this case the Coulomb effects are also significant.

The numerical results for the Delbr\"{u}ck scattering
amplitudes obtained with the use of all four methods at different
$\omega$, $\theta$ and $Z$ can be found in ref.~\cite{8}.
In our work we use the results obtained by method III, which is applicable
under conditions of our experiment.

In the experimental investigations of Delbr\"{u}ck scattering carried out
earlier,  three different photon sources have been used~:

1) Photons from radioactive sources, for instance $^{24} \rm Mg$
($\omega=2.75$ MeV) \cite{9,10}.

2) Photons from nuclear reactions like capture of thermal neutrons in the
energy range  $\omega=4\div12$~MeV~\cite{11,12}.

3) In the energy range $20\div100$~MeV the experiment has been carried out
with tagged bremsstrahlung photons ~\cite{13}. In experiment~\cite{14}
Delbr\"{u}ck scattering above 1~GeV has been investigated using
bremsstrahlung photons without tagging.

The accuracy of photon scattering cross section measurements in the
experiments~[10--13] allowed one not only to observe
Delbr\"{u}ck scattering, but to establish the importance  of the Coulomb
corrections to the cross section as well. Unfortunately, in this energy
range ($\omega\sim m$) the Coulomb corrections are not
calculated up to now, that does not permit to perform the detailed
comparison between  experiment and theory.

At $\omega\gg m$ the amplitudes of Delbr\"{u}ck scattering are calculated
including Coulomb corrections. However, the accuracy of
cross section measurements achieved in experiments [14-15]
is essentially smaller than the accuracy of the theoretical predictions.

In the present experiment tagged backscattered Compton photons
of energies $140\div 450$~MeV  were used. The final photons were
detected at  scattering angles $2.6\div16.6$~mrad. At the specified
parameters the theoretical
predictions based on the quasiclassical approach (method III) have high
accuracy, errors
of the calculation less than one percent. At the same time the 
experimental technique used has
allowed us to get high accuracy of cross section measurements and,
as a result, to make detailed comparison between theory and experiment.

Apart from Delbr\"uck scattering, we have investigated
the process of photon splitting, which has been observed
for the first time. Data analysis and comparison with the accurate theoretical
predictions for this process are in  progress now, preliminary results can be
found in~\cite{15}. Here we only note that the experimental data
within the experimental accuracy ($\sim10\%$) are in agreement  with the
results of recent theoretical calculations performed exactly in $Z\alpha$
\cite{16}.

\section{\bf Experimental setup}

The experiment has been carried out using ROKK-1M facility \cite{17}
of VEPP-4M collider \cite{18}. The experimental setup is shown in
Fig.~2. It includes the solid-state laser, optical system of laser beam
injection to the vacuum chamber of collider, and photon
tagging system \cite{19} of the detector KEDR~\cite{20}.

High energy photons are produced as a result of laser photons backward Compton
scattering on the electron beam.
Scattered photons move along the electron beam and have  a
narrow angular spread of the order $1/ \gamma$, where
$\gamma$ is the electron relativistic factor. The photon energy
spectrum has a sharp edge with the maximum energy, determined by the
following expression~:
\[ E_{max}=\frac{4 \omega_{\scriptscriptstyle L} \gamma^2}{1+4
\omega_{\scriptscriptstyle L} \gamma/mc^{2}}, \]
 where $\omega_ {\scriptscriptstyle L}$ is the laser photon energy
(1.17~eV). The energy of electron beam in experiment was 5.25~GeV, and 
$E_{max}=451$~MeV.

Scattered electron is removed from the beam by bending magnets of
accelerator and falls on the drift tubes hodoscope of the tagging
system. The tagging system allows one to measure the energy of scattered
photon  with the accuracy about 1.3\%.
The minimum energy of tagged photons in the experiment was 140~MeV.

The clean photon beam of small cross sizes is formed by the system of
three collimators. 
The first lead collimator has the 10~cm thickness, $4\times4$~mm$^2$ aperture,
and it actually
sets the cross sizes of the beam. Charged particles of beam halo after this
collimator are removed by
a sweeping magnet with the magnetic field of 550~G. 
The second lead collimator, having the 30~cm thickness and 25~mm the
diameter of a hole, is placed at 13~m from the first one. It absorbs secondary
particles scattered on  large angles. Behind this collimator the third  active
collimator is placed. It is formed by four BGO crystals of the cross sizes
of $25\times25$~mm$^2$ and 150~mm length. An aperture of this collimator
($9\times9$~mm$^2$) was chosen so that its inner edges could not see the
interaction point of the laser photon beam and electron beam.  Scintillation
light from each BGO crystal of this collimator is detected by the
photomultiplier, which signal is included in the trigger of the experiment
in anticoincidence.  The active collimator removes  the impurity of
secondary particles moving in small angles with respect to the main
photon beam. 

The target is placed directly behind the last collimator.  It is BGO
crystal of 12~mm thickness (1.07 radiation length) and $25\times25$~mm$^2$
cross sizes. Light from the crystal is detected by two photomultipliers,
which signals are included in a trigger in anticoincidence. It
enables us  to suppress effectively a background from inelastic
processes in the target at the trigger level. The
threshold for the energy deposition in the target was set at
the level of 150~KeV. 

In order to reduce a background from  Compton scattering after the
target, photons pass the distance between the target and particle detector
(4.8~m) in helium-filled tube of 30~cm diameter.  The photon passed target
without interaction falls onto an absorber installed before the aperture
window of the detector.  The absorber is the BGO crystal cylinder of 23~mm
diameter and 146~mm length.  Its signal is also included in a trigger in
anticoincidence. It permits one to exclude the trigger start up from the
photons, passed target without interaction.  To suppress the trigger
start up from the charged particles,  a thin scintillation veto-counter is
mounted on the aperture window of the detector.

The signals from the BGO-collimator, target, absorber, and veto-counter
are digitized and
used in off-line analysis for background suppression. Besides, the
signals from the target and absorber are used to monitor the input
gamma flux. The typical rate of the initial photons in
experiment was $5-10$~kHz.  These signals are exploited also to form an
additional trigger to start the readout of the tagging system for
measurement of the energy spectrum of initial photons (Fig.3).

Final photons are detected by means of the electromagnetic
calorimeter based on liquid krypton~\cite{21}. The layout of the
calorimeter is shown
in Fig.~4.

The electrodes are made of G10 sheets covered with a copper
foil on the both sides. The first (counting from the input window of the
calorimeter) and all the odd electrodes are under zero potential ("grounded"
electrodes), high voltage is applied to all the even electrodes.

The high-voltage electrodes are used for energy measurements. They are
divided into the 9 pads centered around the beam axis. In the longitudinal
direction the electrode system is segmented into three sections,
altogether the calorimeter contains 27 cells for energy measurements.

The "grounded" electrodes, from second to fourth, are used for the coordinate
measurements. Strips are present on the both sides of these electrodes
and oriented perpendicularly one to another. It enables us to measure
both transverse coordinates (X and Y) in one layer. The strip width varies
from 1~cm in the centre to 3~cm on the edge of the  calorimeter's
aperture. The structure of high-voltage and coordinate electrodes is
shown in details in Fig.~5.   For such coordinate structure of the
calorimeter, the space resolution is equal to 1~mm in the centre and
2.3~mm on the edge. 

The high energy resolution of the calorimeter
($2.4\% / \sqrt{E\makebox[2.7em][l]{(}}$\hspace{-2.2em}GeV)) and photon
tagging allow to use  effectively the equality of the initial and final
photon energies in Delbr\"{u}ck scattering in order
to suppress the background from inelastic processes.

On the initial step of the off-line analysis we select the events
 (named below as events with a detected photon) satisfying
the following criteria:

 1) One track in the tagging system is found.

 2) Energy deposition in the crystals of the BGO-collimator and in the
absorber is less than $0.35$~MeV.

  3) Energy deposition in the target is less than $0.15$~MeV.

 4) Energy deposition in the scintillation counter on the aperture window of
the calorimeter is less than $0.4$~MeV.

 5) Energy deposition in the central tower of the first layer of the
calorimeter is more than $80$~MeV.

 6) One photon in the calorimeter is detected.

In Fig.~6 the example of the event with the detected photon is shown. In this
event photon conversion has happened in the first X-layer.

The efficiency of photon detection  in the calorimeter is about 70\% and
weakly depends on the photon energy at $\omega>140$~MeV  (Fig.~7).
Experimental data shown in Fig. ~7 were obtained in runs when the target and
absorber were removed. It is seen from this picture that the experimental
efficiency is slightly higher than in the simulation. Therefore, in the
comparison between the simulation and experiment this
difference was taken into account.

\section {\bf Background processes for Delbr\"{u}ck scattering}

The total cross section of Delbr\"{u}ck scattering at
$\omega\gg m$ is independent of the photon energy,  and the main contribution
to it comes from scattering angles  $\theta\sim m/\omega$.
In the experiment the photons with the scattering angles
$2.6\div16.6$~mrad were detected.  For bismuth (Z=83 ) the visible
cross section of Delbr\"{u}ck scattering in this range of angles is
equal to $5.9$~mb at $\omega=140$~MeV and $1.2$~mb at
$\omega=450$~MeV. For comparison, the total cross section for bismuth equals
6.4~mb. The contribution to the visible cross section from the 
scattering on germanium (Z=32) is about 3\%.

Compton scattering on atomic electrons is the main background process
under conditions of our experiment. Its cross section for a bismuth
is about $5.3$~mb and
$4.9$~mb at $\omega=140$~MeV and $450$~MeV, respectively.
The results of calculations for differential cross sections of
Delbr\"{u}ck and Compton scattering on bismuth are shown in Fig.~8.

Secondary photons from showers, arising in target and air, give very 
small contribution to the background. The low level of this background is 
the result of the following: the target in the experiment was "active",
and the balance between the energy deposition in the calorimeter 
and the energy of the initial photon was required.

As simulation shows, there is also a small contribution to the
background from photons passed the target and the absorber without
interaction and, in consequence of shower fluctuations in the calorimeter,
detected at $\theta>2.6$~mrad.
The simulation  was made with the use of GEANT  package (version 3.21).

Besides of Compton scattering and electromagnetic cascades
(processes, included in GEANT), photon splitting was also
taken into account as a background process. Since the cross
section of this process is small in
comparison with the cross section of the investigated effect, the
approximate formula (the accuracy $\sim 20\%$) based on the
Weizs\"{a}cker-Williams method was used for the calculation.

Another processes, such as Compton scattering on nuclei,
$\pi^0$ photoproduction, pair production with radiation of a photon, have
been estimated under the conditions of the experiment. The first two
processes give negligible contribution to the number of
scattered photons  detected, while the contribution from the third
one is about $0.5\%$.

Special attention was paid to the background, which comes
from Compton and Delbr\"{u}ck scattering on the edges of the last
BGO-collimator. This question is discussed in Section~IV.

\section{\bf Experimental results and conclusion}

In order to  understand the  situation with background, data has been
taken in two modes: with target and without target (
$9\cdot10^8$ and $2\cdot10^8$ photons, respectively).

In Fig.~9a-d the two-dimensional distributions of energy deposition $E_C$ in
the calorimeter  versus initial photon energy $E_{TS}$  for these modes are
shown. The events with detected photon having the scattering angle
in the range $2.6\div16.6$~mrad are selected.

The diagonal band in Fig.~9 a,c  corresponds to elastic
scattering. In simulation (Fig.~9c) this band contains the events of
Compton scattering and events, in which the initial
photon passed  the target and absorber without interaction and was
detected in the 
given range of angles. The contribution to this band from showers in the
target and air  is very small. In the experimental distribution (Fig.~9a)
this band is enhanced  by virtue of Delbr\"{u}ck scattering in the
target.

It is seen from Fig.~9  that in the experiment there is a large number
of "inelastic scattering" events,  i.~e. events with the energy deposition
in the calorimeter essentially smaller than the energy of the initial
photon. These events are the result of electromagnetic showers
generated in the BGO-collimator by the initial photons touched it.
Since  this effect strongly depends on a position and orientation of the
collimator with respect to the axis  of the photon beam, it is not possible
 to simulate this effect accurately. Nevertheless, to subtract this
background, one can use the experimental data of empty-target runs.
In order to find the correct method of such subtraction, the simulation for
empty-target runs was made.  In the simulation the BGO-collimator was shifted
with respect to the axis of the  initial photon beam so that $0.2\%$ of
photons touched it, and the signal from the collimator was not taken into
account.

Fig.~10 shows the distribution over the parameter $(E_C-E_{TS})/
\sigma_c$  (where $\sigma_c$ is the energy resolution of the calorimeter )
obtained as a result of such simulation. One can see that
the contribution to  the elastic scattering of secondary photons from showers
(generated in the BGO-collimator mainly), when $\Delta
E=|E_C-E_{TS}|<2.5\sigma_c$, is rather small.  The small difference between
simulation and experiment at $\Delta E<2.5\sigma_c$ can be explained
by the fact that Compton scattering in the
BGO-collimator can not be simulated correctly, and Delbr\"{u}ck
scattering in the collimator was not included in
simulation in this case.

Thus, the background from the BGO-collimator for runs with target can be
obtained as a difference between experiment and simulation for empty-target
runs multiplied by the probability for photon to pass target without
interaction. This suppression factor is equal approximately to 0.5 and
weakly depends on the energy at $\omega=140\div450$~MeV.  This is correct at
$\Delta E<2.5\sigma_c$, while for events with $\Delta E>2.5\sigma_c$,
coming mainly from secondary photons, the suppression factor is bigger due
to the charged component of the shower.  It is illustrated by Fig.~11, which
shows, for runs with the target, the distribution  of scattered photons  over
the parameter $(E_C-E_{TS})/ \sigma_c$ with taking this background into
account.

In Table~I we make the comparison between the experiment and the simulation for
different energy ranges of initial photon. The results of simulation 
for Delbr\"uck scattering and the background processes are shown also.
Errors in this Table include statistical errors as well as the total
systematic error. 
The latter one includes the error of measurement of the
initial photons intensity  and the error of photon
detection  efficiency (see Fig.~6). These  errors are estimated to be
$1.5\%$ and $1\%$, respectively.

The differential cross section of Delbr\"{u}ck scattering for
unpolarized photons is given by~\cite{3}
$$\frac{d\sigma}{d\Omega}=(Z\alpha)^4r_0^2\{|A^{++}|^2+|A^{+-}|^2\},$$
where $r_0$ is the classical electron radius,
$A^{++}$ and $A^{+-}$ are non-helicity-flip and helicity-flip amplitudes.

At $\omega\gg m$ and $\Delta\sim m$, as it was in the experiment,
the helicity amplitudes have the form $A\sim\omega f(\Delta)$.
Since $d\sigma / d\Omega$ is independent of azimuth angle $\varphi$,
the differential cross section $d\sigma/dt$ is  equal to
$(\pi/ \omega^2)d\sigma/d\Omega$ (where $t=\Delta^2$) and
depends on the momentum transfer $\Delta$, but not on the
photon energy $\omega$.  It allows one to get the
distribution $d\sigma/dt$ with the use of data for all energies of
the initial photon (Table.~II and Fig.~12).

It is seen from the Table II and Fig.12, that the experimental
results  are in a good agreement
with the theoretical predictions  within the experimental accuracy.

\section*{\bf Acknowledgments}
We express our gratitude to A.N.~Skrinsky and V.A.~Sidorov
for their interest and support of the present research. We
are grateful  to V.P.~Smakhtin for helpful suggestions, and 
V.N.~Baier, A.E.~Bondar, A.P.~Onuchin for useful discussion.
We also thank the experimental staff  of the storage ring VEPP-4M
and technical group of the ROKK-1M facility for providing 
high-quality photon beam. Partial support by Russian Foundation for
Basic   Research (grant RFBR-97-02-18556) is also gratefully acknowledged.

\widetext


\begin{center}Figure captions \end{center}

Fig.~1. (a) Feynman diagrams for  Delbr\"{u}ck scattering:
the Farry representation and the representation via the usual diagrams of
the perturbation theory.
The double line denotes the electron Green function in the Coulomb field,
crosses denote the Coulomb field.
(b) Feynman diagrams in the Farry representation for photon splitting.

Fig.~2. The experimental setup.

Fig.~3. Energy spectrum of the incoming tagged photons.

Fig. ~4. Layout of the calorimeter:  external vessel (1), copper
shield (2), internal vessel (3), signal cable (4),  veto-electrode (5),
strip electrodes (6), towers (L1, L2, L3).

Fig.~5. Electrode structure of the calorimeter.

Fig. ~6. The energy deposition on the strips for the event with the detected
photon (the reconstructed coordinates are X=1.46~cm, Y=--2.56~cm).

Fig.~7. The efficiency of photon detection: experiment (circles),
simulation (line).

Fig.~8. The differential cross section for Delbr\"{u}ck (solid lines)
 and Compton (dashed lines) scattering on bismuth.

Fig. ~9. Distributions of energy deposition $E_C$ in the calorimeter versus
the initial photon energy $E_ { TS }$ for events with the  detected
photons having scattering angles in the range $2.6\div16.6$~mrad.
In simulation only processes included in GEANT are taken into account.
Normalization corresponds to the number of initial photons $1.5\cdot10^8$.

Fig.~10. Distributions of photons,  detected
in the range of angles $2.6\div16.6$~mrad for empty-target runs, over the
parameter $(E_C-E_{TS}) / \sigma_c$. (a) Experimental data (circles) and
simulation (histogram, processes included in GEANT are only taken into account).
(b) Contributions of different effects to the histogram (a):
 Compton scattering (solid line), secondary photons from
showers (dashed line),  photons passed the target and absorber without
interaction (dotted line).  Normalization corresponds to the number
of initial photons $10^9$.

Fig.~11. Distributions of photons,  detected
in the range of angles $2.6\div16.6$~mrad for runs with target, over the
parameter $(E_C-E_{TS})/\sigma_c$. (a) Experimental data (circles) and
simulation with the use of GEANT package (histogram) including Delbr\"{u}ck
scattering, photon splitting, and the background from the BGO-collimator.
(b)  Experimental data (circles), simulation (solid line),
 Delbr\"{u}ck scattering (dashed line), and the sum of all
background processes (dotted line).  Normalization corresponds to the
number of initial photons $10^9$.

Fig.~12. The differential cross section of photon scattering $d\sigma/ dt$
as a function of the momentum transfer $\Delta$
for a molecule of bismuth germanate.
(a) Experimental data (circles) and background (squares). (b)
Experimental data after  subtraction of background (circles).
Solid line is the result of calculation.

\newpage

\footnotesize
TABLE~I. Number of photons, detected  in the range of angles
$2.6\div16.6$~mrad at  $\Delta E<2.5\sigma_c$. Data in each column
are normalized to the number of initial photons $10^9$.

\small
\begin{center}
\begin{tabular}{|c|r@{$\pm$}l|r@{$\pm$}l|r@{$\pm$}l|r@{$\pm$}l|} \hline
\raisebox{0ex}[3ex][2ex]{$\omega$, MeV} & \multicolumn{2}{|c|}{$140\div 450$} &
\multicolumn{2}{|c|}{$140\div 250$} &  \multicolumn{2}{|c|}{$250\div 350$} &
\multicolumn{2}{|c|}{$350\div 450$} \\ \hline
\begin{tabular}{c}Number\\ of initial photons\end{tabular} & \multicolumn{2}{|c|}{902.2$\cdot10^6$} &
\multicolumn{2}{|c|}{ 275.4$\cdot10^6$} &  \multicolumn{2}{|c|}{259.9$\cdot10^6$} & \multicolumn{2}{|c|}{366.9$\cdot10^6$}\\ \hline
  \raisebox{0ex}[3ex][2ex]{Experiment} &
13172 & 232 & 16810 & 353 & 13252 & 301 &10383 &229 \\ \hline
  \raisebox{0ex}[3ex][2ex]{Simulation} &
12810 & 181 & 16709 & 329 & 12535 &  283 &10079 & 209 \\ \hline
 \begin{tabular}{c}Delbr\"{u}ck\\scattering\end{tabular} &
9120 & 111 & 12346 & 171 & 8884 & 150 & 6867 & 119 \\ \hline
  \begin{tabular}{c}Compton\\scattering\end{tabular} &
1334 & 45 & 1624 &  85 & 1254&80 & 1173&66 \\ \hline
  \begin{tabular}{c}Secondary\\photons\end{tabular} &
52&8 & 60&16 & 57&16 & 42&13  \\\hline
  \begin{tabular}{c}Photons passed\\ without interaction\end{tabular} &
495&62 & 954&133 & 324&65 & 270&51  \\\hline
  \begin{tabular}{c}Photon\\splitting\end{tabular} &
435&88 & 434&90 & 519&108 & 376&78  \\\hline
  \begin{tabular}{c}Background from \\the BGO-collimator\end{tabular} &
1374&124 & 1290&246 & 1497&214 & 1351&147\\ \hline
\end{tabular}
\end{center}
\normalsize

\footnotesize
Table II. The differential cross section of Delbr\"{u}ck scattering
$d\sigma/dt$ for a molecule of bismuth germanate. 

\normalsize
\vspace{.5em}
\footnotesize
\begin{center}
\begin{tabular}{|c|c|@{\hspace{2.5\arraycolsep}}r@{$\pm$}l|} \hline
  \raisebox{0ex}[3ex][2ex]{$\Delta$ (MeV)} &
  \begin{tabular}{c} $\displaystyle\frac{d\sigma}{dt}$ (mb/MeV$^2$),\\Calculation\end{tabular} &
 \multicolumn{2}{c|}{\hspace{-2.5\arraycolsep} \begin{tabular}{c} $\displaystyle\frac{d\sigma}{dt}$ (mb/MeV$^2$),\\Experiment\end{tabular}} \\\hline
0.49 & 21.1 & 19.9 & 3.4 \\\hline
0.62 & 12.9 & 10.1 & 1.2 \\ \hline
0.75 & 8.30 & 8.10 & 0.62\\ \hline
0.88 & 5.55 & 5.25 & 0.36\\ \hline
1.01 & 3.83 & 3.88 & 0.25\\ \hline
1.14 & 2.72 & 3.07 & 0.18\\ \hline
1.27 & 1.99 & 2.09 & 0.13\\ \hline
1.40 & 1.49 & 1.66 & 0.11\\ \hline
1.53 & 1.13 & 1.12 & 0.08\\ \hline
1.66 & 0.883 & 0.966 & 0.073\\ \hline
1.79 & 0.696  & 0.758 & 0.064\\ \hline
1.92 & 0.555 & 0.569 & 0.059\\ \hline
2.05 & 0.450 & 0.476 & 0.056\\ \hline
2.18 & 0.369 & 0.410 & 0.055\\\hline
2.31 & 0.306 & 0.356 & 0.045\\\hline
2.44 & 0.255 & 0.238 & 0.049\\\hline
2.57 & 0.215 & 0.148 & 0.038\\\hline
2.70 & 0.183 & 0.139 & 0.035\\\hline
2.83 & 0.157 & 0.187 & 0.035\\\hline
2.96 & 0.135 & 0.151 & 0.028\\\hline
3.09 & 0.116 & 0.112 & 0.030\\\hline
3.22 & 0.101 & (9.65 & 2.7)$\cdot10^{-2}$\\ \hline
3.35 & 8.85$\cdot10^{-2}$ & 0.111 & 0.029\\\hline
3.48 & 7.76$\cdot10^{-2}$ & (9.0 & 2.6)$\cdot10^{-2}$\\\hline
3.61 & 6.83$\cdot10^{-2}$ & (7.0 & 2.3)$\cdot10^{-2}$\\\hline
3.74 & 6.07$\cdot10^{-2}$ & (6.7 & 2.2)$\cdot10^{-2}$\\\hline
3.87 & 5.37$\cdot10^{-2}$ & (6.1 & 2.1)$\cdot10^{-2}$\\\hline
4.00 & 4.80$\cdot10^{-2}$ & (7.1 & 2.0)$\cdot10^{-2}$\\\hline
4.13 & 4.28$\cdot10^{-2}$ & (3.9 & 2.1)$\cdot10^{-2}$\\\hline
4.26 & 3.84$\cdot10^{-2}$ & (3.0 & 1.9)$\cdot10^{-2}$\\\hline
4.39 & 3.46$\cdot10^{-2}$ & (5.0 & 2.1)$\cdot10^{-2}$\\\hline
4.52 & 3.11$\cdot10^{-2}$ & (3.1 & 1.9)$\cdot10^{-2}$\\\hline
4.65 & 2.82$\cdot10^{-2}$ & (4.7 & 1.8)$\cdot10^{-2}$\\\hline
4.78 & 2.53$\cdot10^{-2}$ & (5.3 & 1.7)$\cdot10^{-2}$\\\hline
4.91 & 2.32$\cdot10^{-2}$ & (2.2 & 1.7)$\cdot10^{-2}$\\\hline
5.04 & 2.11$\cdot10^{-2}$ & (3.2 & 1.7)$\cdot10^{-2}$\\\hline
\end{tabular}
\end{center}
\normalsize

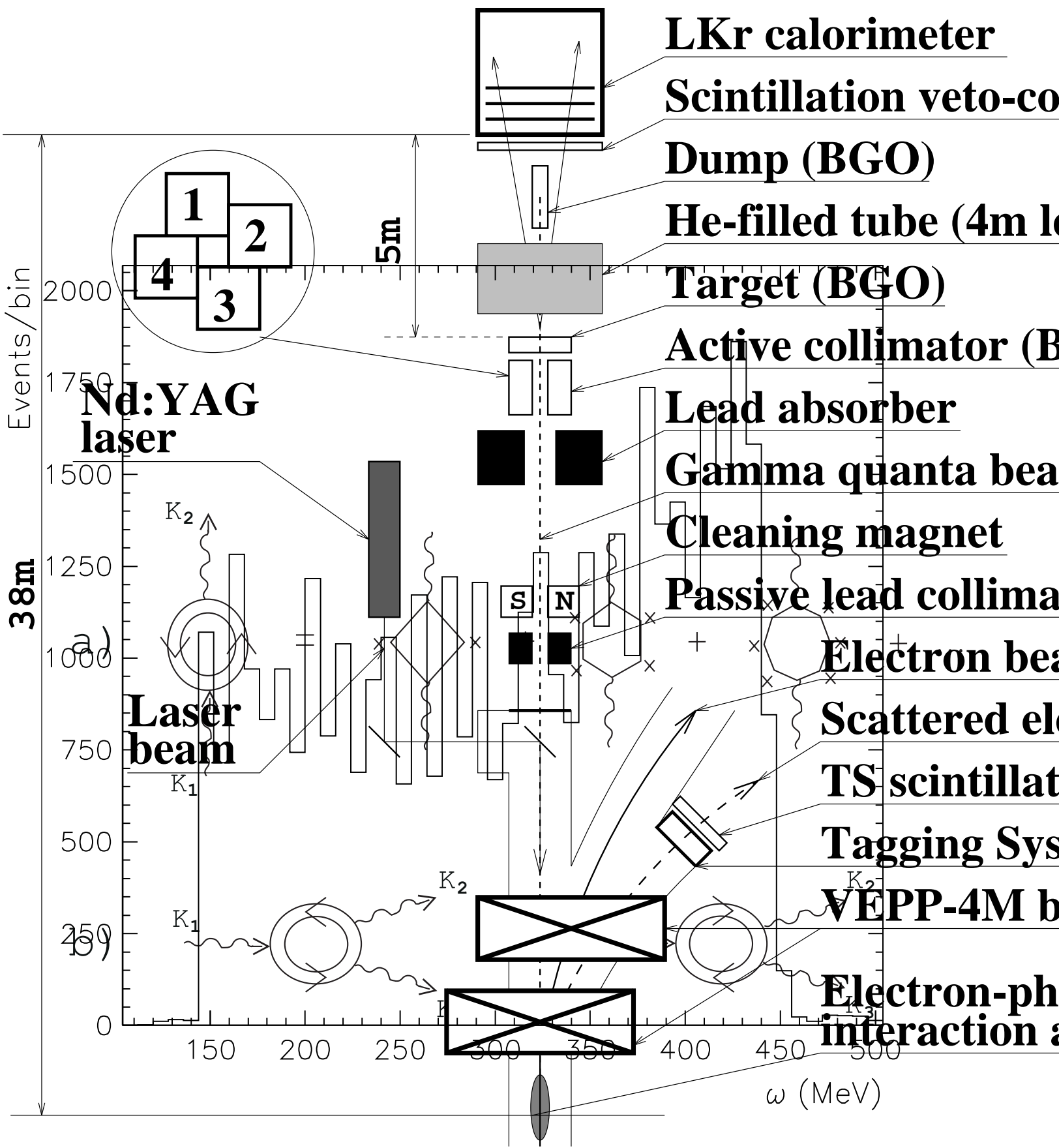
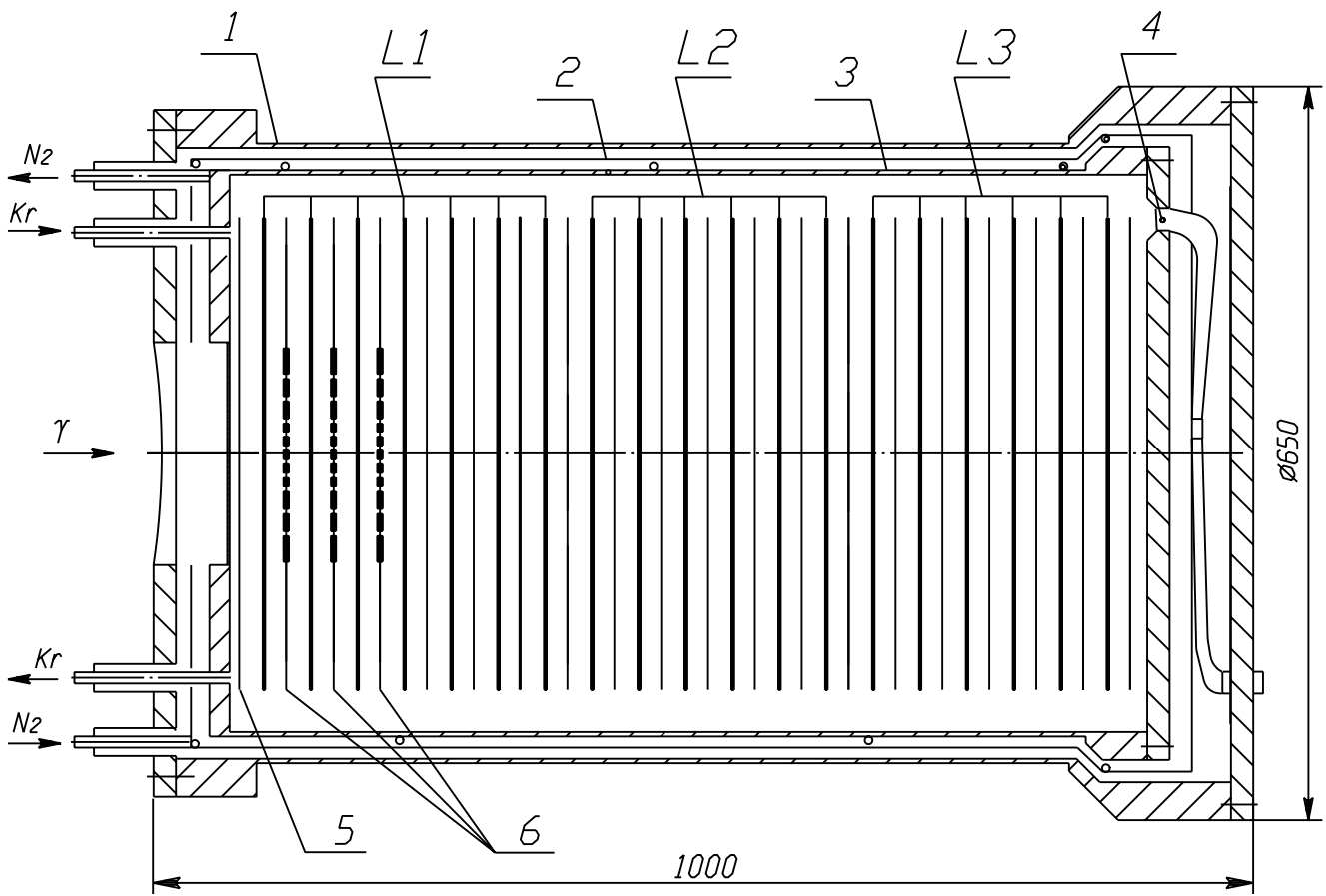
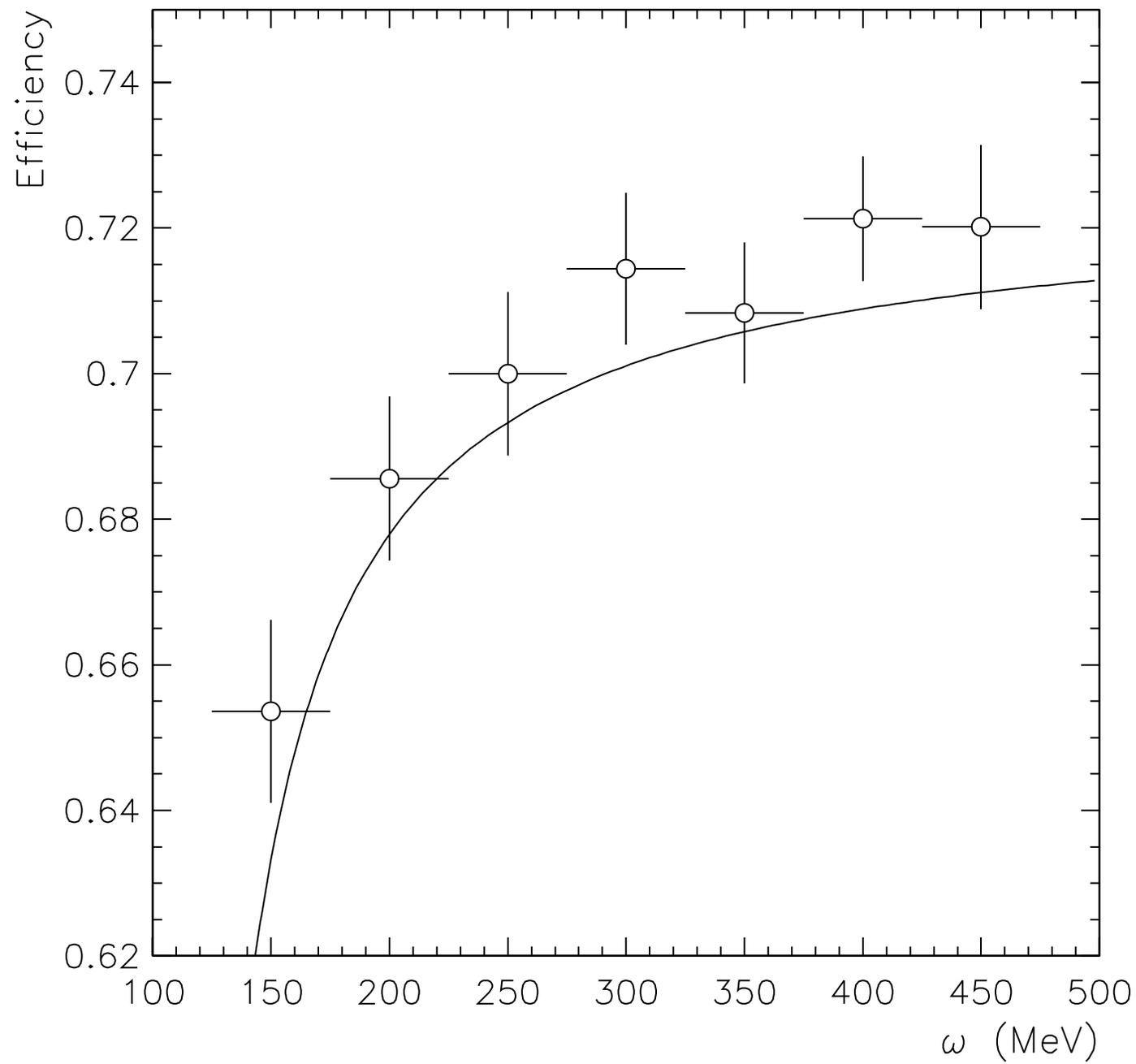
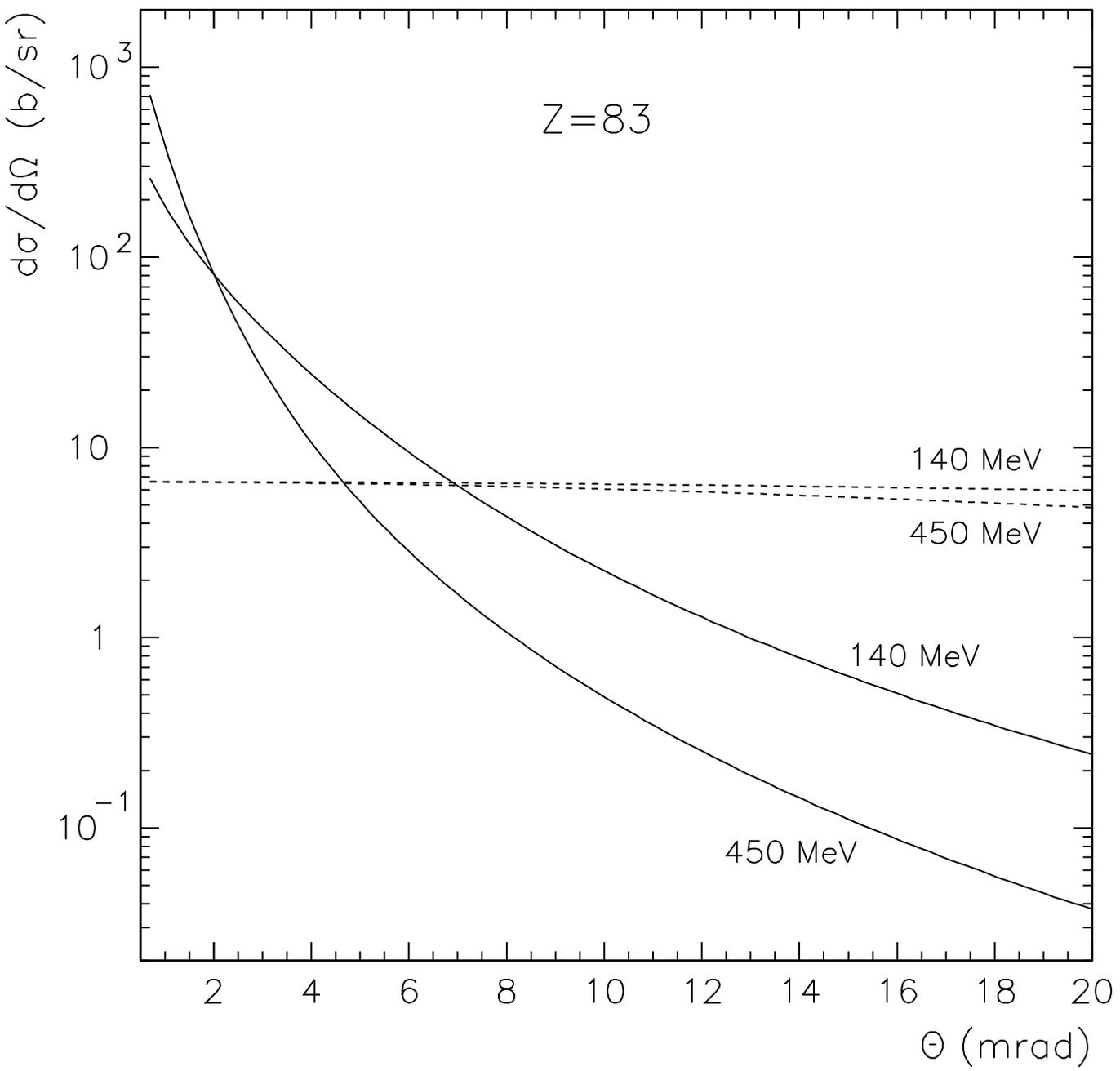
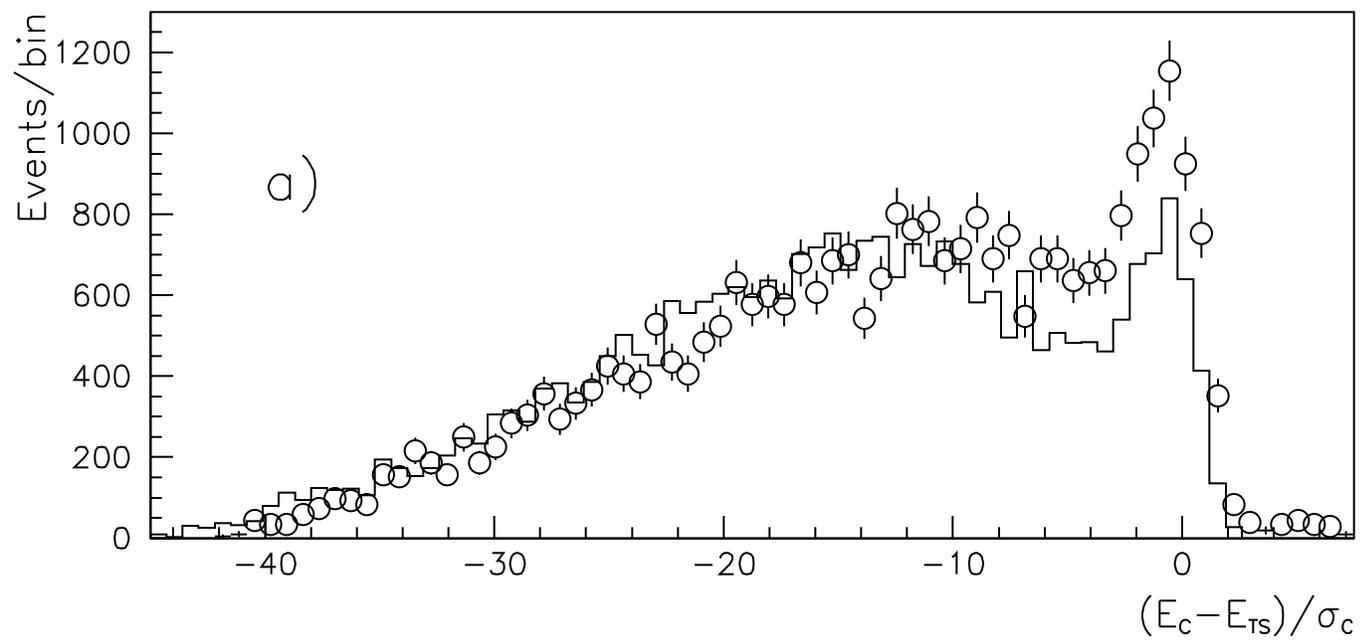
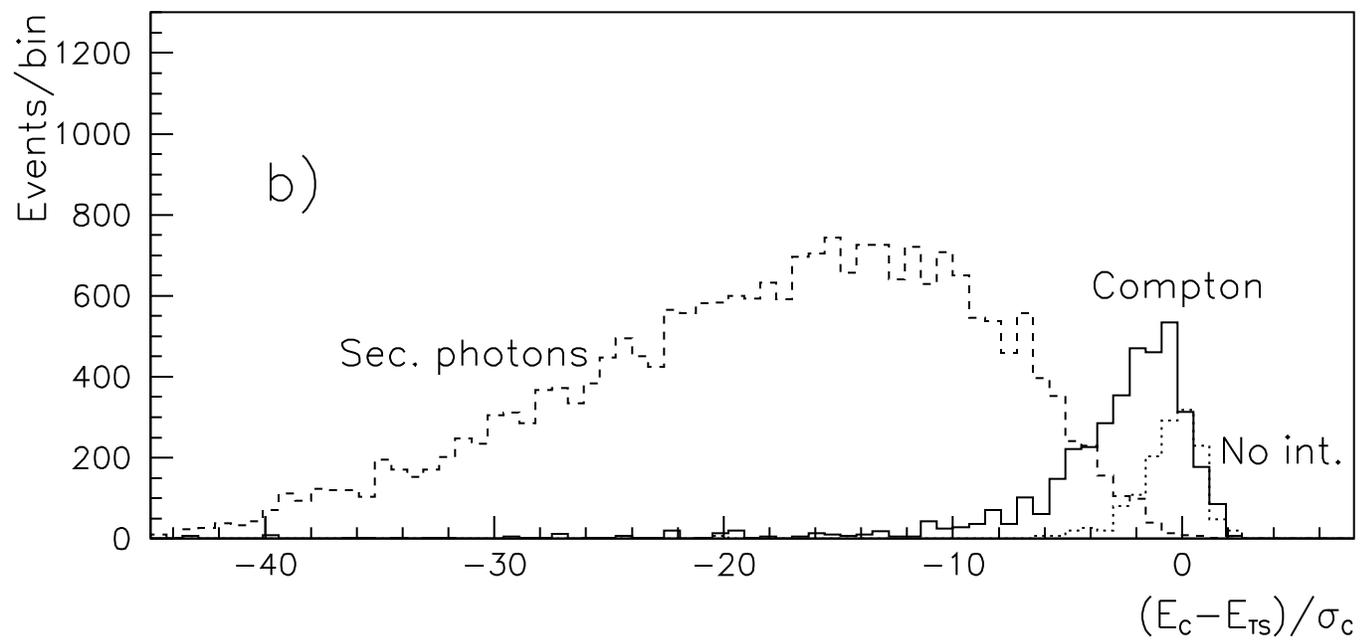
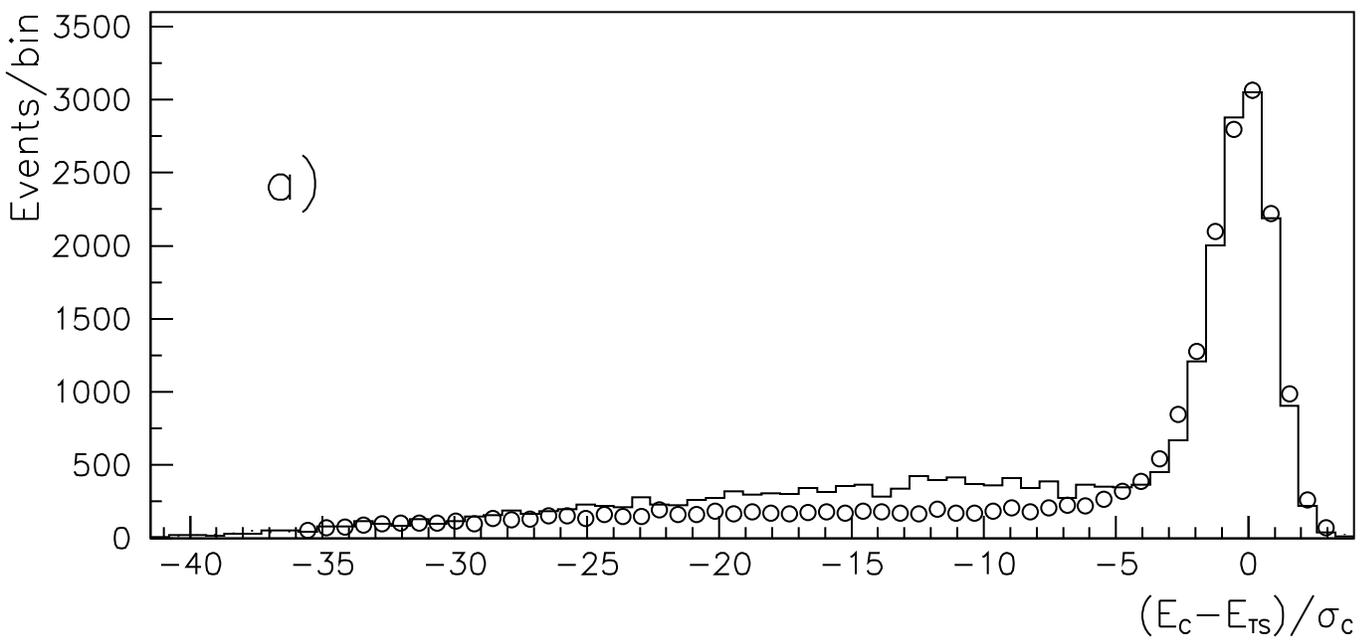
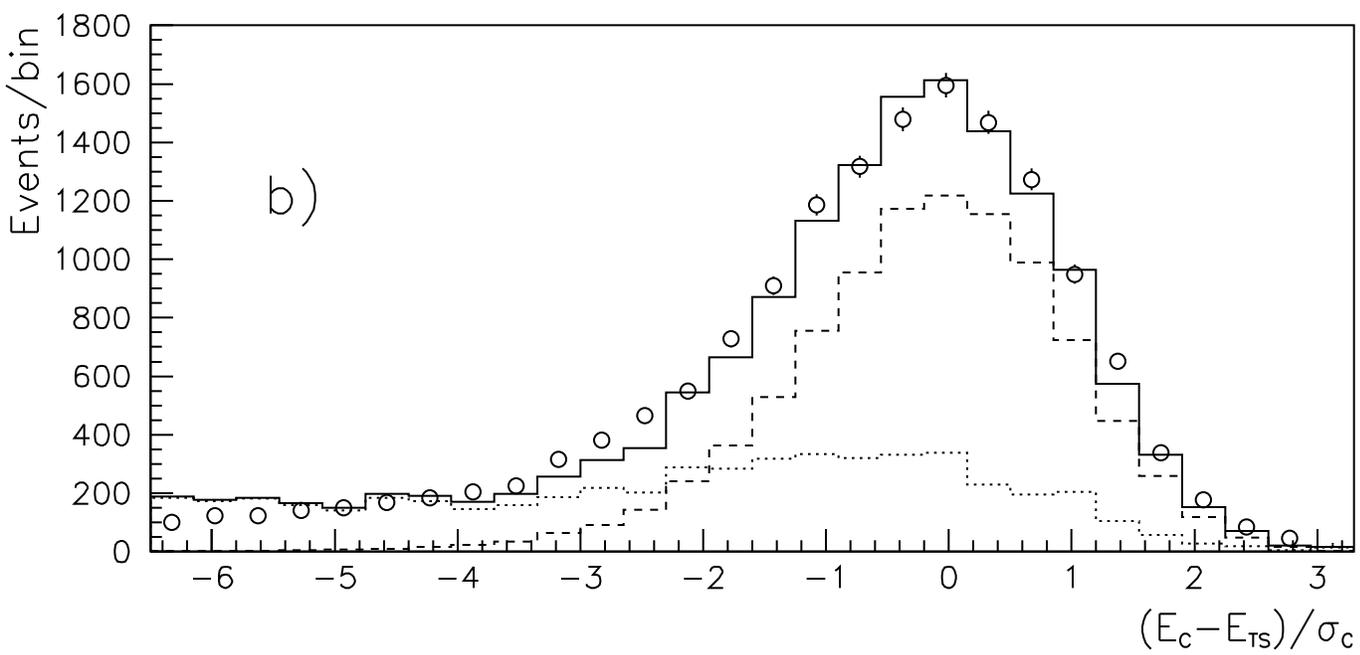
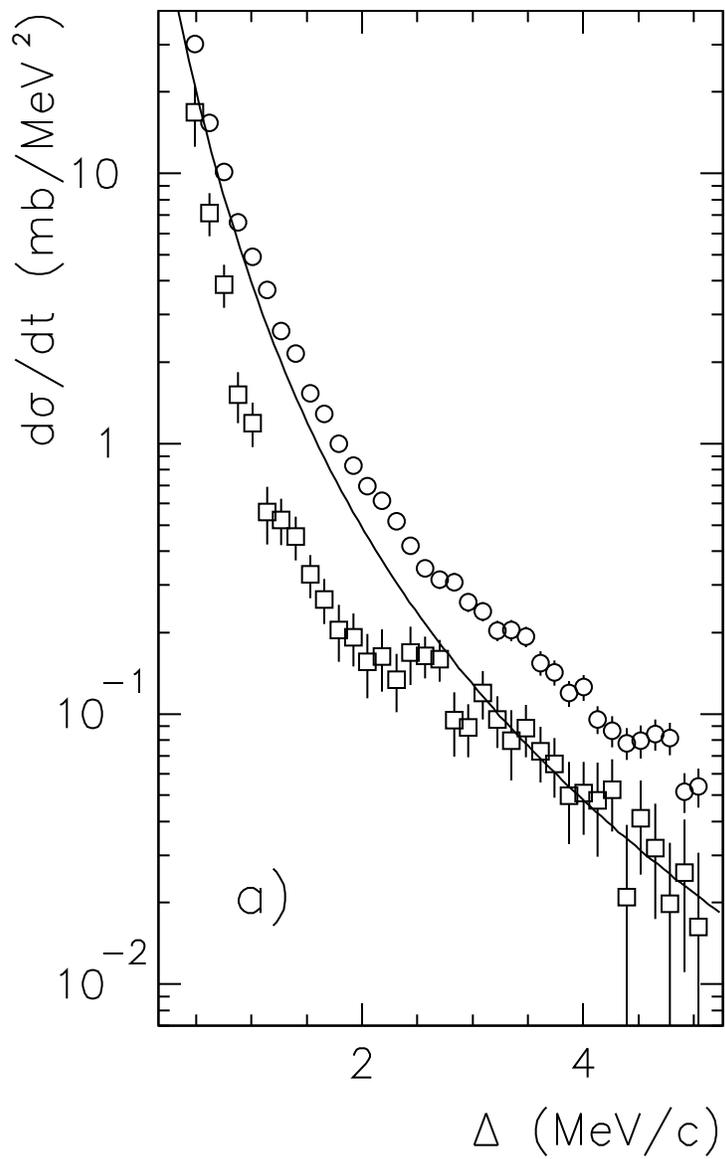
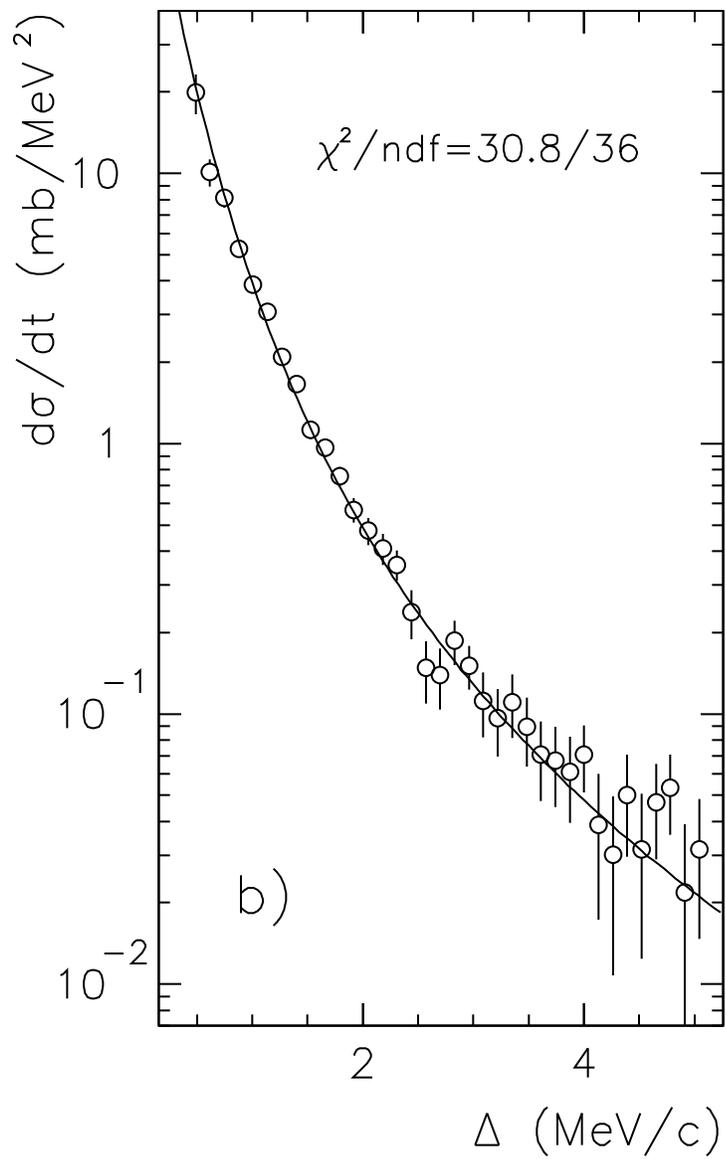
\end{document}